\documentstyle[prl,aps,twocolumn]{revtex}
\large  

\input epsf
\begin{document}
\draft \twocolumn[\hsize\textwidth
\columnwidth\hsize\csname
@twocolumnfalse\endcsname
\title{Ground-state properties of artificial bosonic atoms, Bose interaction 
blockade and the single-atom pipette}
\author{Eugene B. Kolomeisky $^{(1)}$, Joseph P. Straley $^{(2)}$, 
and Ryan M. Kalas  $^{(1)}$} 
\address{$^{(1)}$ Department of Physics, University of Virginia, 382 
McCormick Rd., P. O. Box 400714, Charlottesville, VA 22904-4714\\
$^{(2)}$ Department of Physics and Astronomy, University of Kentucky,
Lexington, KY 40506}
\maketitle
\begin{abstract}
We analyze the ground-state properties of an artificial atom made out
of repulsive bosons attracted to a center for the case that all the 
interactions are short-ranged.  Such bosonic atoms could be created by 
optically trapping ultracold particles of alkali vapors; 
we present the theory describing how their properties depend on
experimentally adjustable strength of ``nuclear'' attraction and 
interparticle repulsion.  The binding ability of the short-range potential
increases with space dimensionality - only a limited number of particles can
be bound in one dimension, while in two and three dimensions the
number of bound bosons can be chosen at will. Particularly in three
dimensions we find an unusual effect of enhanced resonant binding:
for not very strong interparticle repulsion the equilibrium number of
bosons bound to a nuclear potential having a sufficiently shallow 
single-particle state increases without bound as the nuclear
potential becomes {\it less attractive}.  As a consequence of the competing
nuclear attraction enhanced by the Bose statistics and interparticle 
repulsions, the dependence of the ground-state energy of the atom on
the number of particles has a minimum whose position is experimentally 
tunable.  This implies a staircase dependence of the equilibrium
number of bound bosons on external parameters which  may be used to
create a single-atom pipette - an arrangement which allows the transport of 
atoms into and out of a reservoir, one at a time.
\end{abstract}
\vspace{2mm} 
]

\narrowtext

\section{Introduction}

The experimental observation of Bose-Einstein condensation (BEC) of
trapped alkali vapors \cite{ander} has triggered a tremendous volume
of experimental and theoretical work in this field
\cite{Anglin&Ketterle}.  One of the attractive features of BECs which
makes them an active area of research is that they represent
well-controlled ensembles of atoms that can be used to study novel
aspects of many-body physics. 

In this work we will describe an effect and its device implementation
which will farther improve control over the number of particles in a trap:  
we explain how to build a quantum pipette enabling single-atom
manipulation.  The operation of the pipette is based on the discrete
nature of the particles combined with their mutual repulsion.  The
pipette is a bosonic cousin of the single-electron box
\cite{Likharev}, however the differing statistics (Bose versus Fermi)
and the nature of the interactions (short-range repulsion among bosons 
versus long-range Coulomb repulsion between the electrons) require a 
separate analysis.      

A way to create a quantum tweezer for atoms was put forward recently 
\cite{Raizen}.  Even though the underlying physics is related to what
follows, the proposed extraction scheme cannot work.   It was claimed that by 
moving an attractive potential out of the BEC at certain speeds, a chosen 
number of atoms could be extracted by means of Landau-Zener tunneling 
\cite{LL1}.   It was assumed that the superfluidity of the BEC helps 
prevent excitations which otherwise would be induced by the motion;
this restricts the speeds to be smaller than the sound velocity.
However the local sound velocity {\it vanishes} at the edge of the 
condensate \cite{DGP}.  Thus it is impossible to exit the BEC at
finite velocity without incurring nonsuperfluid effects.  Moreover 
Ref.\cite{Raizen} relies on a model which, as will be explained below, cannot 
be the starting point of a realistic computation. 

We call our arrangement a pipette because it allows both uploading and
downloading at the single-particle level.  As in Ref.\cite{Raizen} the
role of the pipette 
is played by an external short-range attractive potential which can be created 
at the focus of a red-detuned laser beam \cite{Leggett}.  In order to 
understand the mechanism by which single-particle manipulation becomes 
possible, we need to know the ground-state energy $E_{n}$ of $n$
repulsive bosons bound by the potential.  

To set the conventions, select the zero of the external potential at infinity 
and assume that it only has one single-particle state of energy 
$E_{1} < 0$.   For noninteracting bosons, all $n$ particles will
populate this state thus leading to a ground-state energy that
decreases linearly with the number of 
particles:  $E_{n} = -|E_{1}|n$.  The effect of interparticle repulsion may 
be accounted for heuristically by noticing that there are $n(n-1)/2$ pair 
interactions among $n$ particles.  Then the ground-state energy can be 
estimated as $E_{n} = -|E_{1}|n + \nu n(n-1)/2$, where 
${\nu} > 0$ is proportional to the amplitude of the interparticle repulsion. 

Imagine that the pipette is placed in the vicinity of a BEC reservoir formed 
by a potential which does not vary significantly on the spatial scale of the 
pipette, as shown in Fig.1.  Then the overlap of the reservoir and pipette
potentials shifts the single-particle level of the pipette to 
$E_{1} + U_{res}({\bf r}_{p})$ 
where the reservoir potential $U_{res}({\bf r})$ is evaluated at the 
location of the pipette.  The ground-state energy calculated relative to 
the chemical potential $\mu$ of the reservoir, 
$E_{n}^{'} = E_{n} + [U_{res}({\bf r}_{p}) - \mu]n$, is given by \cite{Raizen}
\begin{equation}
\label{hgsenergy}
E_{n}^{'} = [-|E_{1}| + U_{res}({\bf r}_{p}) - \mu]n + \nu n(n-1)/2
\end{equation}
Eq.(\ref{hgsenergy}) overlooks some important aspects of the interacting 
problem to be discussed below, but it correctly captures the basic 
physics of the effect, and will be used to motivate our study.  Since the 
pipette is outside the BEC [$U_{res}({\bf r}_{p}) > \mu$], they are 
separated by a tunneling barrier whose strength can be tuned by
varying the distance between the pipette and the reservoir.  
The particles of the reservoir will still be able to tunnel into the 
pipette, and there can be particles in the pipette 
whenever the chemical potential of the BEC ($\mu$) is higher than the 
single-particle level [$E_{1} + U_{res}({\bf r}_{p})$] shown in Fig.1.
So for the particle extraction to occur in the presence of the
tunneling barrier, the dimensionless parameter 
$\gamma = [|E_{1}| + \mu - U_{res}({\bf r}_{p})]/|E_{1}|$ should lie 
between zero and unity. 
\begin{figure}[htbp]
\epsfxsize=3.6in
\vspace*{-0.3cm}
\hspace*{-0.5cm}
\epsfbox{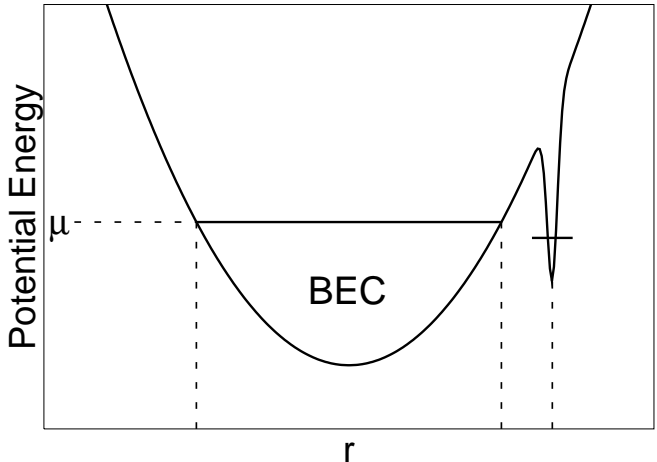}
\vspace*{0.1cm}
\caption{Potential energy landscape where the smooth
external potential confining a BEC, $U_{res}({\bf r})$, overlaps
with that of a sharp short-range well (the pipette) as a function of 
position.  The pipette is placed some distance away from the BEC edge,
$U_{res}({\bf r}) = \mu$, so that a tunneling barrier with the
condensate exists.  The particles can be transported in and out of the 
pipette one at a time by tuning both the depth of the short-range
potential and the strength of interparticle repulsion.  The short
horizontal segment inside the pipette indicates its single-particle state.}
\end{figure}
When $\gamma$ is close to unity, the pipette is nearly inside the condensate.
The coupling to the reservoir is strong, so that the particles can easily
tunnel in and out of the pipette.  Under these conditions
the pipette will not be in a state of definite particle number:  it will
be in a quantum superposition state, with amplitudes for different
integer pipette populations.  This means that the particle discreteness is
irrelevant, and as far as expectation values are concerned,  $n$ in 
(\ref{hgsenergy}) can be treated as a continuous variable.  The 
average number of bound bosons is then given by the minimum of 
(\ref{hgsenergy}), which occurs for $<n> = n_{x} = 1/2 + |E_{1}|\gamma/\nu$.  We note
that the parameter $n_{x}$ can be controlled experimentally:  the 
single-particle energy $E_{1}$ can be varied by changing the depth of
the pipette potential which in turn can be accomplished by adjusting
the power of the focused laser beam.  For fixed pipette position this
also affects the tunneling barrier.  The strength of the interparticle 
repulsion $\nu$ can also be tuned by applying an external magnetic
field which alters the proximity of the Feschbach resonance
\cite{Leggett}.  We will use $n_{x}$ as the independent variable in
what follows to represent these ways of controlling the system.

The opposite limit of weak coupling to the reservoir will be our main 
interest, because then single-atom manipulation is possible.
Although the external parameter $n_{x} = 1/2 + |E_{1}|\gamma/\nu$ 
is generally not an integer, the pipette population will nearly
always take on fixed integer values \cite{note1}.  
Half-integer values of $n_{x}$ play a special role because now two
states of the pipette with $n = n_{x} \pm 1/2$ bosons have the same 
energy.  Tunnelling mixes these two states of the system so that as $n_{x}$ is 
slowly varied, the $n$ particle state adiabatically evolves
into an $n \pm 1$ particle state.  When the tunnelling is small, 
the increase or decrease in particle number occurs over a small range
of $n_{x}$, so that the change seems almost discontinuous.  In what follows we
will refer to these changes as population transitions keeping in mind
that the notion of the transition is strictly valid only in the limit
of zero tunneling and correspondingly infinitesimally slow variation of
$n_{x}$.
\begin{figure}[htbp]
\epsfxsize=3.7in
\vspace*{-0.3cm}
\hspace*{-0.7cm}
\epsfbox{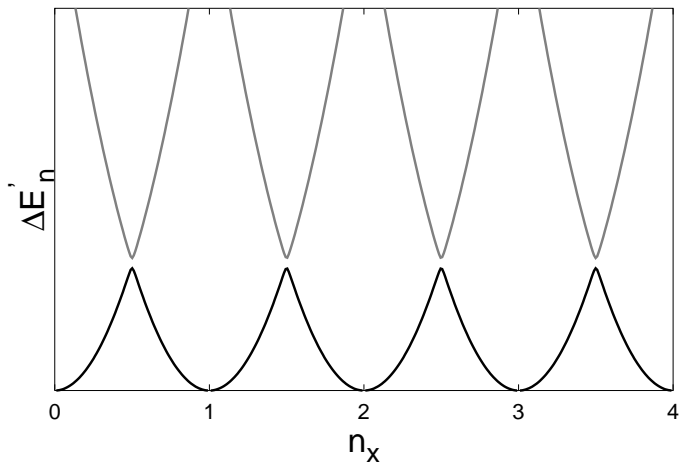}
\vspace*{0.1cm}
\caption{The $n_{x}$ dependence of the ground-state
energy $\Delta E_{n}^{'}$ for a series of integer $n$ for the model
(\ref{hgsenergy}).  The lower bold curve corresponds to the
ground-state.  Gray scale is used for metastable states.}
\end{figure}        
To help visualize the population transitions it is useful to look
at the ground-state energy computed relative to its minimum, 
$\Delta E_{n}^{'} = E_{n}^{'} - E_{n_{x}}^{'} = \nu(n - n_{x})^{2}/2$.
The loci of the population transitions are then given by 
$\Delta E_{n}^{'} = \Delta
E_{n - 1}^{'}$, and the corresponding graphical construction is shown in
Fig.2 where we plot the $n_{x}$ dependence of $\Delta E_{n}^{'}$ for
a series of integer $n$.  This would give us a series of intersecting
parabolas, each corresponding to a fixed number of 
bound bosons given by its minimum.  Similar to the electronic case
\cite{Likharev}, mixing of the states of different 
particle number replaces the intersections of the energy curves 
at half-integer values of $n_{x}$ by small energy gaps as shown in
Fig.2.  As the external parameter $n_{x}$ changes adiabatically, the
system follows the lowest energy curve shown in bold.  In the limit of
zero tunneling the switches between different parabolic segments can
be regarded as first-order population transitions because the slopes
of the intersecting energy curves do not coincide.
      
As the barrier between the pipette and reservoir changes from barely 
penetrable to almost transparent, the equilibrium staircase 
$<n> = [n_{x}]$ evolves into a linear dependence $<n> = n_{x}$ as
sketched in Fig.3.  We note that a finite BEC temperature also
contributes into smearing of the dependence of $<n>$ on $n_{x}$.   
\begin{figure}[htbp]
\epsfxsize=3.6in
\vspace*{-0.3cm}
\hspace*{-0.5cm}
\epsfbox{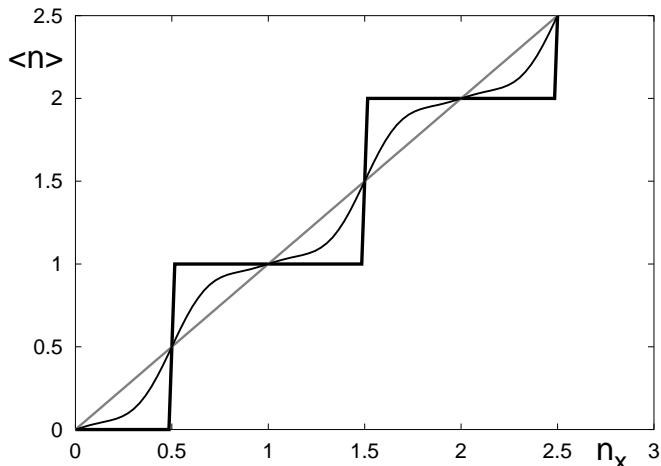}
\vspace*{0.1cm}
\caption{The $n_{x}$ dependence of the equilibrium population of the
pipette $<n>$ for different levels of transparency of the barrier with
the BEC:  barely penetrable (bold solid staircase), almost transparent
(gray $<n> = n_{x}$ line), and in between (thin solid curve).}
\end{figure}  
The role of a finite tunneling rate is twofold:
First it leads to the rounding of staircase dependence shown in Fig. 3:  
the discontinuities at degeneracy points are replaced
by rapid variations near half-integer $n_{x}$.  Second, it constrains
the rate of change of the external parameter $n_{x}$ to be slower than
the tunneling rate to allow for the single-particle extraction to take 
place.  Otherwise the system will not follow the lowest energy path
shown in Fig. 2:  the dependence of $<n>$ on $n_{x}$ 
will exhibit metastability 
and hysteresis.

Here is an example of the pipette operation.  Imagine turning
on the laser beam adiabatically, focussed at a fixed place {\it
outside} the BEC, but not very far from its edge, as depicted in Fig.1.  
As the laser 
power increases, so does the amplitude of the pipette potential, and
when it reaches a critical depth, a single-particle bound state 
appears \cite{LL2}.  Further increase of the power moves this
(yet unpopulated) state down in energy until it reaches the BEC chemical 
potential $\mu$ (see Fig.1).  At that point the states with zero and
one particles in the pipette become degenerate, 
$E_{0}^{'} = E_{1}^{'}$, and upon infinitesimal increase of the power
a particle tunnels into the pipette.  As 
the laser power increases further, the pipette remains singly occupied until 
the energy gain of having two extracted particles 
overwhelms the penalty paid in increase of interparticle repulsion:
the second particle enters the pipette at 
$E_{1}^{'} = E_{2}^{'}$.  A further increase of the pipette population in 
unit increments happens at special values of the laser power as it 
goes up:  single-particle tunneling out of the BEC reservoir occurs 
whenever the degeneracy condition $E_{n}^{'} = E_{n-1}^{'}$ is met.
The blocking of single-particle tunneling by repulsive interactions 
among bosons -- the Bose interaction blockade -- parallels the phenomenon
of the Coulomb blockade in electronic systems \cite{Likharev}. 

After a desired number of particles has been extracted, the focus of the laser 
can be translated nonadiabatically to somewhere else \cite{translate},
and then the pipette can be emptied, releasing the stored
particles one by one by adiabatically decreasing the beam power.   To
verify that the correct number of atoms has been extracted,
the atoms could be released into a magneto-optical trap (MOT).
Atoms in a MOT scatter light at a well-defined rate, so by measuring the
scattered power small numbers of atoms can be accurately counted
\cite{Hu94,Ruschewitz96}.

Let us now critically assess the physical content of the model.  
The functional form (\ref{hgsenergy}) tells us that $n$ 
bosons are condensed into the ground state of the pipette potential and that
$n(n-1)/2$ pair interactions are distributed over a spatial scale which 
purports to be independent of the number of particles $n$.  This scale can
only be the localization length of the single-particle ground state.  
Therefore the size of the many-body state corresponding to 
(\ref{hgsenergy}) is the same as that of the single-particle state:  the model
(\ref{hgsenergy}) only attempts to account for the increase of energy due to 
interparticle repulsion but overlooks the simultaneous ``swelling'' of the 
ground-state.  As the number of bound particles increases, so does the 
localization length of the ground state.  This in turn decreases the 
zero-point energy per particle thus effectively introducing an $n$
dependence into $E_{1}$.  At the same time, the interparticle
repulsions are weakened by being distributed over a length scale which 
grows with $n$.  We conclude that the model of interacting particles 
populating a single-particle state [as described by Eq.(\ref {hgsenergy})]
is valid only for extremely weak interparticle repulsion 
[$\nu(n-1)/|E_{1}| \ll 1$], when the swelling of the ground state can 
be neglected.  With this in mind, the conclusion that $E_{n}^{'}$ has 
a minimum is actually beyond the range of applicability of (\ref{hgsenergy}).
Below we will exhibit a series of counterexamples to the functional form
(\ref{hgsenergy}) including the practically important case of a 
three-dimensional system.  

Our goal now is to compute $E_{n}^{'}$ in a controlled fashion and convince the
reader that parts of the physics based on Eq.(\ref{hgsenergy}) are in
fact correct.  We start from the more fundamental problem of
calculating the many-body ground-state energy $E_{n}$ in the presence
of a confining potential that vanishes at infinity.  In view of the
recent achievement of an all-optical BEC \cite{Chapman} where the 
trapping potential is short-ranged, solving this problem has an importance of
its own.  The ground-state energy $E_{n}^{'}$ calculated relative to the 
BEC chemical potential and accounting for the overlap with the 
reservoir potential then simply follows as 
$E_{n}^{'} = E_{n} + [U_{res}({\bf r}_{p}) - \mu]n = E_{n} + (1 - \gamma)|E_{1}|n$.

The rest of the paper is organized as follows.  In Section II we
introduce the self-consistent (Hartree-Fock) approximation which is our
main calculational tool and pose the problem using a language
resembling the Coulomb terminology of atomic physics.  In Section III
a one-dimensional problem is studied where we start by discussing an
exact minimizer of the Hartree-Fock theory (Section IIIA).  In Section IIIB we 
consider a model of repulsive bosons on a half-line attracted to the 
origin.  Although this problem represents a considerable deviation from that
treated in Section IIIA, it has a benefit of exact solvability thus 
allowing us to clarify certain aspects of the self-consistent treatment.   
Higher-dimensional cases are investigated in Sections IV 
(between one and two dimensions) and V (two dimensions) where
variational solutions to the Hartree-Fock theory are given. A
variational solution for a three-dimensional problem is
described in Section VI.  Not only is this case the practically most
relevant, but it is also somewhat unusual in terms of physics:  we find
that there is a range of parameters when the number of bound particles 
can {\it increase} as the trapping potential gets {\it less attractive}.  
In Section VII we present a Landau-Zener \cite{LL1} type theory which explains
tunneling smearing of the boson staircase sketched in Fig.3.  In 
Section VIII we discuss feasibility of experimental implementation
of the single-particle manipulation.  Since the single-particle
effects are enhanced in traps of small range, first (Section VIIIA) we
describe a method of creating an optical trap which is considerably
tighter than those in current use.  We conclude (Sections VIIIB and
VIIIC) by discussing experimental limitations set by nonzero
temperature of the condensate and finite time of tunneling. 

\section{Formulation of the problem}

A controlled analytical calculation of the ground-state properties of a 
system of $n$ repulsive bosons attracted to a center $U({\bf r})$ largely 
avoiding model assumptions is possible in the limit that the interaction 
range of the trapping potential $a$ is significantly smaller than the 
localization length $R$ of the ground state.  Assume in addition that the 
$d$-dimensional trapping potential is radially symmetric, $U({\bf r}) = U(r)$, 
and decays sufficiently rapidly so that 
$S_{d} V = - \int U( {\bf r} ) d^{d}r 
= - \int \limits _{0}^{\infty}U(r) S_{d} r^{d - 1}dr > 0$ is finite.  Then the 
condition $R \gg a$ implies that $n$ bosons are effectively bound to an 
attractive delta-function center of strength $S_{d}V$ where $S_{d}$ is
the area of the $d$-dimensional unit sphere. 

We call this system an artificial bosonic ``atom'' (with the pipette 
potential playing a role of the nucleus), since the inequality 
$R \gg a$ parallels the situation in the physics of real atoms,  
where the size of the nucleus is significantly smaller than the atomic 
size.  In contrast to the atom of fermions wherein the 
Pauli principle dictates most of the properties and interactions play 
secondary role, the physics of the bosonic atom is dominated by interparticle
repulsion \cite{chbose}, and even a nucleus with only one shallow 
single-particle state may bind more than one particle.

For general $n$ and $d$ the many-body properties cannot be computed exactly 
in a system with broken translational symmetry, and we have to resort to 
approximate methods.  Then the permutation symmetry of
the problem implies that the ground-state wave function can be
approximated by the Hartree-Fock product ansatz:  $\Psi({\bf
r}_{1},...,{\bf r}_{n}) = \prod \limits _{i = 1}^{n}(\psi({\bf
r}_{i})/\sqrt{n})$ where the ``single-particle'' wave function 
$\psi({\bf r})$ minimizes the energy functional:    
\begin{equation}
\label{GP}
E=\int d^{d}x \left [ {\hbar ^{2}\over 2m} (\nabla \psi)^{2}
+ {g (n-1)\over 2n}\psi ^{4} \right ] \ - S_{d}V\psi^{2}(0)
\end{equation}
Here $m$ is the particle mass, $g$ [proportional to $\nu$ in 
(\ref{hgsenergy})] accounts for the short-range repulsion among the
bosons, and the wave function is normalized so that
$\psi^{2}$ is the particle density, i. e. 
$n = \int d^{d}x \psi^{2}({\bf r})$.  The interaction term of (\ref{GP}) 
tells us that each particle moves in a self-consistent potential provided by 
the remaining $n-1$ particles, and the simplified form of the last term 
describing interaction with the nucleus is due to the condition $R \gg a$. 

For $n=1$ minimization of Eq.(\ref{GP}) is equivalent to the variational 
principle of quantum mechanics while for $n \gg 1$ the $n$-dependence drops out
of (\ref{GP}) and we arrive at the Gross-Pitaevskii theory 
\cite{DGP,GP}.  When substituted back in the Hartree-Fock-Gross-Pitaevskii
(HFGP) energy functional (\ref{GP}), the radially-symmetric wave 
function $\psi({\bf r}) = \psi(r)$ minimizing Eq.(\ref{GP}) provides us 
with the ground-state energy $E_{n}$.  

In what follows we will often use dimensionless parameters
that contain information about various aspects of the problem:
\begin{equation}
\label{parameters}
\xi = {mVa^{2-d}\over \hbar^{2}} \simeq {mU_{0}a^{2}\over
\hbar^{2}},~Z = {S_{d}V \over g},~v = {\xi \over Z},~\lambda = {n - 1 \over
Z}
\end{equation}
The strength of the nuclear attraction is parametrized by
$\xi$, which is the typical depth of the nuclear well
$U_{0}$ measured in units of the zero-point energy of a particle
localized within the spatial scale $a$.  The relative strength of the
nuclear attraction and interparticle repulsion is
characterized by the reduced nuclear ``charge''  $Z$ which in a real atom 
is the number of protons in the nucleus.  The dimensionless ratio 
$v$ parametrizes interparticle repulsions alone.  The parameter
$\lambda$ characterizes the degree of ``ionicity'';  in a standard
atom (ion) it tells us about the relative number of protons and
electrons.  Given that all the particles are neutral, the
Coulomb language is used figuratively. 

Since the binding properties of a short-range potential depend 
on the spatial dimensionality, we will consider the $d = 1, 2$ and 
$3$ cases separately.

\section{One-dimensional case}

In one dimension at distances satisfying $|x| \gg a$ the
single-particle ground-state wave function decays 
as \cite{LL3} $\psi(x) \sim e^{-|x|/R_{1}}$ where 
$R_{1}= \hbar^{2}/2mV$ plays the role of an ``atomic'' unit of 
length.  This behavior can be continued to the origin without introducing a 
divergence which implies that we can set $a=0$ from the beginning. The 
consistency condition $R_{1} \gg a$ reduces to the requirement of a shallow 
well, $\xi \ll 1$. 

\subsection{Self-consistent treatment}

For $a = 0$ the minimization of (\ref{GP}) is mathematically identical
to a problem solved by Lieb and collaborators \cite{Lieb} in the
context of the ground-state properties of an ordinary atom placed in
an extremely strong magnetic 
field.  For general $n$ the normalized wave function 
minimizing (\ref{GP}) is given by 
\cite{Lieb}
\begin{equation}
\label{Liebwf}
\psi(x) = {\sqrt n(2 - \lambda)\over 2\sqrt {R_{1}\lambda}  
\sinh[{2-\lambda\over 2R_{1}}|x|+ C]}
\end{equation}
Here $\tanh C = (2 - \lambda)/\lambda$, and the 
minimum of (\ref{GP}) exists only for $\lambda \le 2$.  For $\lambda = 2$ 
Eq.(\ref{Liebwf}) reduces to 
$\psi(x) = \sqrt{(Z + {1 \over 2})R_{1}}/(|x| + R_{1})$.  The 
localization length of the many-body ground state follows from (\ref{Liebwf}) 
as $R_{n} = R_{1}/(1 - {\lambda\over 2})$. In "atomic" units this becomes
\begin{equation}
\label{lld=1}
\rho_{n} = R_{n}/R_{1}= \left (1 - {n -1 \over 2Z}\right )^{-1}
\end{equation}
This result implies that a given particle of the atom effectively sees 
a nuclear charge $Z$ screened by an amount proportional to the
number of remaining $n - 1$ bosons.

Eqs. (\ref{Liebwf}) and (\ref{lld=1}) show that as the number of bound 
particles increases, so does the size of the atom, eventually diverging at 
$n = n_{max} = 2Z + 1$, which is the maximum number of bosons that can be 
bound.   The latter can be viewed as a critical phenomenon:  for $n = n_{max}$ 
the wave function is no longer well-localized but still normalizable. 

The reduced ground-state energy $\epsilon_{n} = E_{n}/|E_{1}|$, measured in 
``atomic''  units of the single-particle energy 
$|E_{1}| = 2mV^{2}/\hbar^{2}$ \cite{LL3}, is found to be \cite{Lieb}
\begin{equation}
\label{gsenergyd=1}
\epsilon_{n} = - n \left [1 - {n - 1 \over 2Z} + {(n - 1)^{2}\over
12Z^{2}}\right ]
\end{equation}
We note that even at $n = n_{max} = 2Z + 1$ the ground-state energy
remains negative.  
\begin{figure}[htbp]
\epsfxsize=3.6in
\vspace*{-0.3cm}
\hspace*{-0.5cm}
\epsfbox{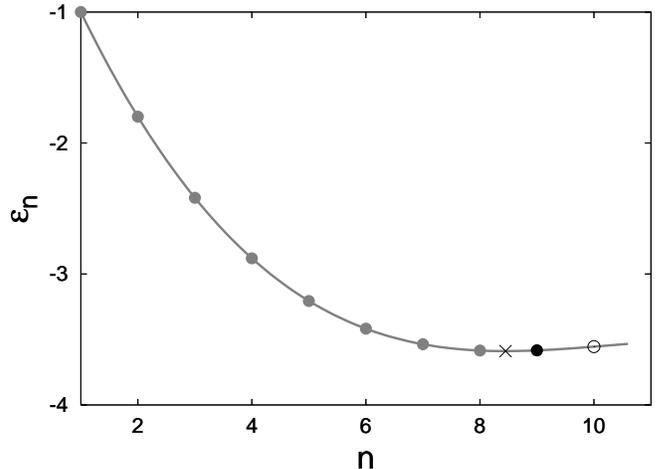}
\vspace*{0.1cm}
\caption{The dependence of the ground-state energy of a
one-dimensional Bose atom on the number of bound particles $n$ for $Z =
4.8$.  The minimum of the curve indicated by the cross is an average
population of the atom strongly coupled to a zero chemical potential
bosonic reservoir.  In the opposite limit of a well-isolated atom the
allowed integer occupation states are shown by the circles - the ground
state (solid circle) and metastable states (gray and open circles).}
\end{figure}

A self-consistent treatment based on (\ref{GP}) is valid in one dimension only 
when the system is sufficiently dense \cite{KNSQ}, so that direct interactions 
are more important than zero-point motion.  For a fixed number of 
bound bosons $n$, however, the size of the atom (\ref{lld=1}) increases 
as $Z$ approaches the critical condition $n = n_{max} = 2Z + 1$:  the 
particle density everywhere decreases.  Here the relevant quantity to
look at is the reduced density 
$\hbar^{2}\psi^{2}(x)/mg = Z(Z + {1\over 2})/({|x|\over R_{1}} + 1)^{2}$ which 
must be significantly larger than unity \cite{KNSQ}.  Applying this condition 
first to the center of the density distribution, we find that 
$\hbar^{2}\psi^{2}(0)/mg \gg 1$ provided 
$Z \gg 1$.  As one moves away from the center, the density 
decreases and at $|x| \simeq ZR_{1} \gg R_{1}$ the parameter 
$\hbar^{2}\psi^{2}(x)/mg$ becomes of order unity.  It is then straightforward 
to verify that in the tails of the 
distribution $|x| \gtrsim ZR_{1}$ there is roughly one particle out of 
$2Z + 1 \gg 1$ total.  Thus for all practical purposes the tail of 
the distribution is irrelevant,  the particles are not farther away from
the center than $ZR_{1}$, and the self-consistent treatment is justified.  
The latter conclusion also holds away from the critical 
condition $n = n_{max}$.  Rigorous analysis \cite{Lieb} concludes that 
the self-consistent treatment based on (\ref{GP}) produces
asymptotically exact results, Eqs.(\ref{Liebwf})-(\ref{gsenergyd=1}), 
for $n, Z \gg 1$ with the ratio $n/Z$ fixed.

Beyond the range of their asymptotic accuracy 
Eqs.(\ref{Liebwf})-(\ref{gsenergyd=1}) can be used as interpolation 
formulas:  for $n = 1$ they automatically reproduce the exact results;
the second term of (\ref{gsenergyd=1}) can be also recognized as a
result of the first-order perturbation theory in $(n - 1)/Z \ll 1$.

The population dependence of the ground-state energy
(\ref{gsenergyd=1}) typical to the case of weakly-repulsive particles (not very
small nuclear charge $Z$) is sketched in Fig.4. 

For $n \le n_{max} = 2Z + 1$ the ground-state energy has a minimum at 
$n = n_{x} = 2Z + {2\over 3} - \sqrt{(6Z + 1)/9}$.  The position 
of this minimum can be externally controlled, and it determines the average 
population of the atom connected through an easily penetrated barrier to
a zero chemical potential bosonic reservoir (see Fig.4).  We note that the
minimum becomes more shallow as $Z$ increases.

In the opposite limit of weak coupling to the reservoir, the particle 
discreteness is important and the atomic population is quantized.  The
equilibrium number of bound bosons is determined by an integer $n$
such that the energy $\epsilon_{n}$ is closest to $\epsilon_{n_{x}}$.
In Fig.4 this is indicated by the bold solid circle.   All other 
integer-valued population states are metastable and shown by gray and
open circles.  We make a distinction between metastable atoms whose
population is smaller (gray circles) or larger (open circles) than the
optimal because the latter can lower their energy by ejecting a
particle while the former cannot. As a result, well-isolated underpopulated
atoms are more stable than their overpopulated counterparts.  In
equilibrium particle exchange takes place when the degeneracy between
the states with $n$ and $n - 1$ particles occurs, 
$\epsilon_{n} = \epsilon_{n-1}$, i. e. at 
$n = n_{1} = 2Z +{7\over 6} - \sqrt{(24Z +1)/36}$ which is a 
first-order transition.  We note that the inequality 
$n_{x} < n_{1} < n_{max}$ always holds. 

We also note that in the limit $Z \rightarrow \infty$ the difference 
$n_{1} - n_{x}$ approaches $1/2$.  This has a simple interpretation:
if one is only interested in the population transitions, then for $Z \gg 1$ the
energy dependence (\ref{gsenergyd=1}) can be approximated by a
parabola centered at $n = n_{x}$.  As a result we arrive at a model
similar to (\ref{hgsenergy}) with half-integer $n_{x}$ playing a special role.

In order to understand the implications of these results to the
operation of the pipette, we need to look at the ground-state energy   
calculated relative to the BEC chemical potential and accounting for
the overlap with the reservoir potential 
$E_{n}^{'} = E_{n} + [U_{res}({\bf r}_{p}) - \mu]n = E_{n} + (1-\gamma )|E_{1}|n$.
Its dimensionless counterpart $\epsilon_{n}^{'} = E_{n}^{'}/|E_{1}| = 
\epsilon_{n} + (1-\gamma ) n$ follows from (\ref{gsenergyd=1})
\begin{equation}
\label{gsenergy'd=1}
\epsilon_{n}^{'} = - n \left [\gamma - {n - 1 \over 2Z} + {(n -
1)^{2}\over 12Z^{2}}\right ]
\end{equation}  
The differences of the binding properties of the pipette from those of
the atom are due to the mismatch of the chemical potential $\mu$ 
and the trapping potential of the BEC at the pipette location
$U_{res}({\bf r}_{p})$.  In the
regime of interest when the particle extraction takes place in the
presence of the tunneling barrier, $0 < \gamma < 1$,  the energy gain
of condensation is diminished.
The analysis of
(\ref{gsenergy'd=1}) is similar to that of 
(\ref{gsenergyd=1}), and leads to similar results.

\subsection{Exactly solvable example}

The self-consistent treatment of the previous Section predicts that 
population transitions are discontinuous.  However for the exactly
solvable case of two point repulsive bosons in the presence of a
delta-function potential well \cite{Rosenthal} a transition from two
to one bound particles is known to be continuous.  Below we obtain an exact
solution of a one-dimensional many-body problem where {\it all}
population transitions are continuous.  In addition to clarifying the
nature of the population transitions this solution provides us with an 
alternative derivation of the  $n \gg 1$ limit of Eqs.(\ref{lld=1})
and (\ref{gsenergyd=1}).  Solving the many-body problem exactly
becomes possible due to a choice of the pipette potential $U(x)$ which 
allows application of the Bethe ansatz methods \cite{Bethe}.

The many-body Hamiltonian has the form
\begin{equation}
\label{Hamiltonian}
\hat{H} = \sum \limits_{i = 1}^{n}\left [-{\hbar^{2}\over 2m}{d^{2}\over d
x_{i}^{2}} + U(x_{i})\right ] + g \sum \limits_{i < j} \delta(x_{i} - x_{j})
\end{equation}
We select the trapping potential $U(x)$ to have an attractive part and
an impenetrable core causing the wave function
of a particle immediately next to it to fall off as
$\exp(-x/R_{1})$ - now the particles are only allowed on the 
half-line $x > 0$.  The potential can be heuristically pictured as a 
delta-function well located right next to an impenetrable wall.
We select the decay length $R_{1}$ as in Section IIIA:  $R_{1}
= \hbar^{2}/2mV$, i. e. as if it were due to a delta-function well
of strength $2V$.  Shortly it will become clear why this choice is reasonable. 

An {\it attractive}, $g < 0$, version of the many-body problem
(\ref{Hamiltonian}) has been solved by Kardar \cite{Kardar} in the
classical context of the line depinning transition in two dimensions
in the presence of quenched impurities.  In pure system the
transfer-matrix technique reduces this problem to the quantum
mechanics of a particle in a trapping potential.  With disorder
present, the replicated problem corresponds to $n$ mutually attracting
bosons in the same trapping potential. The interesting physics
pertinent to the random problem is revealed in the limit of zero
number of replicas, $n \rightarrow 0$ \cite{Kardar}.

Although Kardar focused on attractive interactions \cite{Kardar}, his 
arguments do not depend to the sign of the interparticle
interaction.  Some of the results given below can be deduced from those of
Ref. \cite{Kardar} by simply reversing the sign of interparticle interaction.

For each permutation P of the particles with $0 < x_{P1} < x_{P2} <
... < x_{Pn}$ the bosonic wave function is written as a product of
exponentials:
\begin{equation}
\label{Bethewf}
\Psi \sim \exp\left (-\sum \limits _{i = 1}^{n}\kappa_{i} x_{Pi}\right
)
\end{equation}  
Selecting the ``momenta'' $\kappa_{i}$ to satisfy
\begin{equation}
\label{momenta}
\kappa_{i} = {1\over R_{1}} - {mg(i - 1)\over \hbar^{2}} = {1\over
R_{1}}\left (1 - {i - 1 \over Z}\right )
\end{equation}
will guarantee the correct derivative discontinuities upon two-particle
exchange.  Also for the particle closest to the attractive center the wave
function has the required $\exp(-x_{P1}/R_{1})$ fall-off.  Since the
wave function (\ref{Bethewf}) is an eigenstate of the Hamiltonian
(\ref{Hamiltonian}) with no nodes, it describes the ground state of
the problem. 

Eqs.(\ref{Bethewf}) and (\ref{momenta}) imply that the reduced size of the
ground state is given by
\begin{equation}
\label{Bethell}
\rho_{n} = R_{n}/R_{1} = \left (1 - {n - 1 \over Z}\right )^{-1}
\end{equation}
This result shows that as the number of bound particles increases, so
does the size of the ground state, eventually diverging at $n = n_{max}
= Z + 1$ which is the maximal number of bosons that can be bound.
Similar to Eq.(\ref{lld=1}) screening of the nuclear attraction is the 
mechanism responsible for the divergence of the localization length 
(\ref{Bethell}).

The ground-state energy can be computed as $E_{n} = -\sum \limits_{i
=1}^{n}\hbar^{2}\kappa_{i}^{2}/2m$.  For the reduced ground-state
energy $\epsilon_{n} = E_{n}/|E_{1}|$ measured in units of the
single-particle energy $|E_{1}| = \hbar^{2}/(2mR_{1}^{2}) =
2mV^{2}/\hbar^{2}$ we find
\begin{equation}
\label{Bethegsenergy}
\epsilon_{n} = - n \left [1 - {n - 1\over Z} + {(n - 1)(2n - 1)\over
6Z^{2}} \right ]
\end{equation}
We note that for $n = n_{max} = Z + 1$ the ground-state energy remains
negative.

The problem posed by the Hamiltonian (\ref{Hamiltonian}) can be viewed
as a simplified version of that of a one-dimensional Bose atom treated
in Section IIIA.  Indeed consider two sets of $n/2$ point repulsive
bosons each constrained to occupy positive and negative half-lines and 
attracted to a delta-function center of strength $2V$ at the origin.
Since the sets are noninteracting, this is merely two systems described by the
Hamiltonian (\ref{Hamiltonian}) (with $n$ replaced by $n/2$) put 
``back to back'' to each other.  It is then natural for the wave function
for $x <0$ to be a mirror of the wave function for $x > 0$.  This
explains why it was reasonable to require that the wave function of
the particle next to the origin decays as $\exp(-x/R_{1})$ with $R_{1}
= \hbar^{2}/2mV$ because now we built a discontinuity in the derivative
of the wave function at $x =0$ which is appropriate for a
delta-function of strength $2V$.  The reduced size of the Bose-atom
can be deduced from (\ref{Bethell}) as $\rho_{n} = [1 - (n -
2)/(2Z)]^{-1}$ where we simply replaced $n$ by $n/2$.  This result is
very similar to Eq.(\ref{lld=1}):  they have common $n \gg 1$ limit
and for $Z \gg 1$ predict the same maximal number $2Z$ of
bound bosons.  Similarly, the ground-state energy of the
one-dimensional Bose-atom can be deduced from (\ref{Bethegsenergy}) as
$\epsilon_{n} = - n [1 - (n - 2)/(2Z) + (n -1)(n - 2)/(12Z^{2})]$
where we replaced $n$ by $n/2$ and doubled the outcome since there are
two sets each containing $n/2$ particles.  This result is very similar
to Eq.(\ref{gsenergyd=1}) and they have common $n \gg 1$ limit.  It is
unclear whether these $\rho_{n}$ and $\epsilon_{n}$
improve on Eqs.(\ref{lld=1}) and (\ref{gsenergyd=1}) as the exact
solution to the one-dimensional Bose-atom problem should account for 
arbitrary partitions of $n$ particles.
\begin{figure}[htbp]
\epsfxsize=3.6in
\vspace*{-0.3cm}
\hspace*{-0.5cm}
\epsfbox{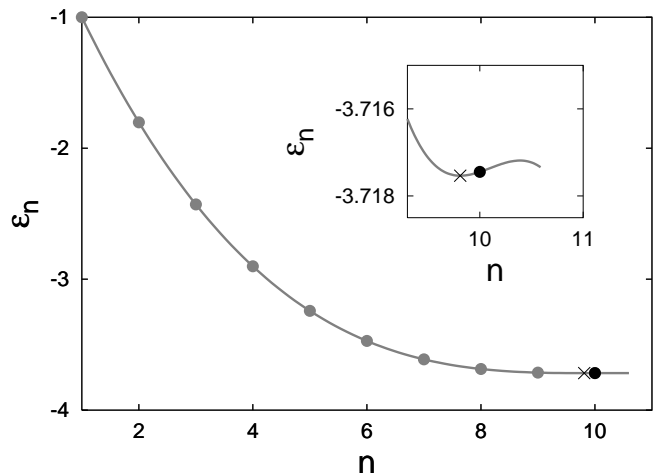}
\vspace*{0.1cm}
\caption{Dependence of the ground-state energy of a one-dimensional
model, Eq.(\ref{Hamiltonian}), on the number of bound particles $n$, 
Eq.(\ref{Bethegsenergy}), for $Z = 9.6$.  The styling is the same as
in Fig.4 and the inset shows the behavior of the energy curve in the
vicinity of its minimum.}
\end{figure}
Typical dependence of the ground-state energy (\ref{Bethegsenergy}) on
the number of bound bosons $n$ is shown in Fig.5.  In producing Fig.5
we intentionally chose $Z = 9.6$ which is twice the value used to draw 
Fig.4 because according to the argument of the previous pararaph this 
allows us to make a comparison with the self-consistent treatment. The 
difference between Figs.4 and 5 is hardly percievable
to the eye and manifests itself in the vicinity of the minimum of the
energy function shown in the inset of Fig.5.   

For $n \le n_{max} = Z + 1$ the ground-state energy
(\ref{Bethegsenergy}) has a minimum at $n = n_{x} = Z + {1\over 2} -
{\sqrt{3}\over 6}$ and a maximum at $n = Z + {1\over 2} +
{\sqrt{3}\over 6}$ with fixed $Z$-independent distance of $\sqrt{3}/3
< 1$ between them.  The position of the minimum, indicated in Fig.5 by
the cross, determines the average number of bound bosons provided the
trap is connected through easily penetrable barrier to a zero chemical 
potential bosonic reservoir.  Similar to the case of the
one-dimensional Bose-atom, Section IIIA, as interparticle
interactions weaken, i. e. $Z$ increases, the minimum becomes
increasingly more shallow.

In the opposite limit of weak coupling to the reservoir the particle
discreteness is important and the number of bound bosons is determined
by an integer $n$ for which the energy $\epsilon_{n}$ is closest to 
$\epsilon_{n_{x}}$.  The particle exchange takes place when the
degeneracy between the states with $n$ and $n - 1$ particles occurs, 
i. e. $\epsilon_{n} = \epsilon_{n-1}$.  It is straightforward to
verify that this condition
{\it coincides} with the condition of divergence of the localization
length (\ref{Bethell}): $n = n_{1} = n_{max} = Z + 1$.  Moreover at
the transition point the derivative of the ground-state energy with
respect to $Z$ is continuous, $\partial \epsilon_{n = Z + 1}/\partial Z =
\partial \epsilon_{n = Z}/\partial Z = - (Z^{2} - 1)/(3Z^{2})$.  These
facts imply that the population transitions are continuous. 
\begin{figure}[htbp]
\epsfxsize=3.7in
\vspace*{-0.3cm}
\hspace*{-0.8cm}
\epsfbox{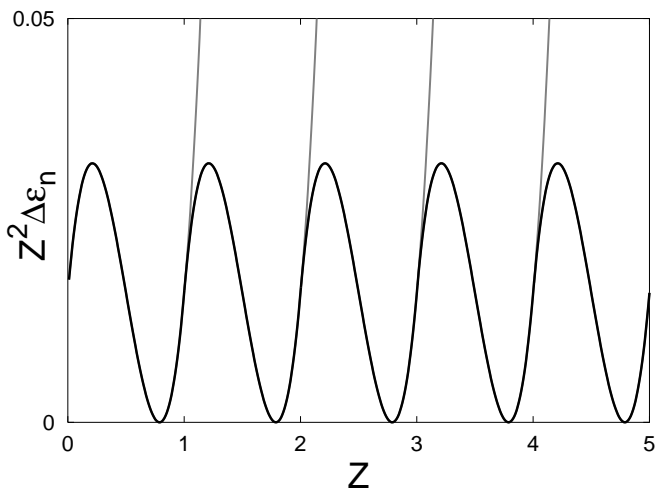}
\vspace*{0.1cm}
\caption{Dependences of $Z^{2}\Delta \epsilon_{n}$ of the model
(\ref{Hamiltonian}) on the interaction parameter $Z$ for a series of
integer $n \le Z + 1$.  The styling is similar to that in Fig.2.  Gray
lines are continuations of $Z^{2}\Delta \epsilon_{n}$ on $Z$ for $Z >
n$, while $Z^{2}\Delta \epsilon_{n}$ does not exist for $Z < n - 1$.}
\end{figure}
The continuity of the population transitions is easier to visualize by
looking at the ground-state energy computed relative to its minimum,
$\Delta \epsilon_{n} = \epsilon_{n} - \epsilon_{n_{x}}$.  The
transitions take place at $\Delta \epsilon_{n} = \Delta
\epsilon_{n - 1}$, and  Fig.6 in a fashion similar to Fig.2 shows the
$Z$ dependences of $Z^{2}\Delta \epsilon_{n}$ for a series of integer
$n \le Z + 1$.  As the interaction parameter $Z$ changes
adiabatically, the system follows the lowest energy path shown in bold.  
Although Fig. 6 resembles Fig. 2, there is some difference in their
origin: in the model (\ref{Hamiltonian}) the states of differing $n$
are not coupled by tunneling. Nonetheless, the switches between neighboring
energy branches are smooth, and the lowest energy state (represented
this way) is a strictly periodic function with unit period.  

In order to understand why for the model (\ref{Hamiltonian}) the
population transitions are continuous, we note that the Hartree-Fock
wave function assumes that the ground state is Bose-condensed - every
particle is assigned the same function.  On the other hand, in the exact 
wavefunction (\ref{Bethewf}) each particle is described by a different
function. As a result the ground-state has a shell-like structure -
there are $n$ localization lengths in the problem.  The particles that
are farther out act like they are bound by a weaker potential screened
by all the particles of inner shells.  This implies that the nuclear
potential is screened more effectively than self-consistent treatment 
would predict.  The way the system evolves from an $n$ particle 
wavefunction to an $n - 1$ particle wavefunction is by making the wave
function of the outermost particle extremely extended in space - only
the radius of the largest shell diverges while all other remain finite.
Therefore at the point of the population transition the ground-state
energy $E_{n}$ merges without discontinuity with that of the $n - 1$
particle wavefunction. 

Since the exactly solvable model (\ref{Hamiltonian}) has basic
ingredients of any model of repulsive bosons in the presence of a center, we
argue that (\ref{Hamiltonian}) defines a universality class:  the
shell structure of the ground-state is generic and screening is the
mechanism leading to continuous population transitions. 

In trying to apply this reasoning to higher-dimensional cases we note
that only in one dimension arranging the repulsive particles in certain
order with respect to the attractive center necessarily lowers the
energy relative to that of the Hartree-Fock state.  In higher
dimensions this does not have to be the case as there is more ``room'' 
around the nucleous and thus the screening is less effective.  As a
result the ground state is most likely to be Bose-condensed, and the 
Hartree-Fock single-shell wave function adequately represents physics.

We note that in the vicinity of the particle reservoir the conditions of 
degeneracy are modified due to the mismatch of the BEC chemical
potential and the trapping potential evaluated at the pipette
location.  As a result the transitions when the pipette population
changes are discontinuous regardless of the nature of the population 
transitions of Bose-atoms.

\section{Variational treatment:  ${\bf 1 \le d < 2}$}

For general dimension $d > 1$ we could not find a closed form
minimizer of the HFGP energy functional (\ref{GP}), and instead used a
variational method.  The idea behind selecting a good trial wave
function is to have it resemble as much as possible the
single-particle wave function.

To prepare the reader for the practically
important two- and three-dimensional cases, and to assess the accuracy of
the variational treatment in general, first we approximate the trial wave 
function by a simple exponential, $\psi(r) \sim e^{-r/R}$, with
$R$ being the variational size of the ground state.  Normalizing it
and substituting in (\ref{GP}) we arrive at the variational energy
\begin{equation}
\label{varenergygend}
E(R) = n \left [{\hbar^{2}\over 2mR^{2}} 
- {V(2^{d + 1} - \lambda) \over 2 \Gamma(d)R^{d}}\right ]
\end{equation}
The first term of
(\ref{varenergygend}) represents the zero-point energy cost of having 
$n$ bosons
localized within the range $R$, while the second term is the
potential energy consisting of energy gain of attraction to the
nucleus and the energy cost of interparticle repulsion.  Since $R$ is the
only scale in the problem, all the $R$ dependences in
(\ref{varenergygend}) can be recovered heuristically from
dimensionality arguments:  apart from numerical coefficients any
function with localization scale $R$ will lead to (\ref{varenergygend}).   

The $E(R)$ dependence  (\ref{varenergygend}) has a minimum provided $0
< d < 2$ and $\lambda < 2^{d +1}$, i. e. for $n < n_{max} = 1 + 2^{d + 1}Z$.
For $n = 1$ minimizing $E(R)$ reproduces the correct answers for the
localization length  $R_{1} = [\hbar^{2}\Gamma(d)/2^{d}mdV]^{1/(2 -
d)}$ and the  ground-state energy $E_{1} = E(R_{1}) 
= - [\hbar^{2}(2 - d)/(2md)] [2^{d}mdV/\hbar^{2} \Gamma(d)]^{2/(2 -
d)}$ \cite{KS};  the consistency condition $R_{1} \gg a$ becomes $\xi \ll 1$.
For general
$n$ minimizing (\ref{varenergygend}) we find the reduced size of
the ground-state
\begin{equation}
\label{varllgend}
\rho_{n} = R_{n}/R_{1}= \left (1 - {n -1 \over 2^{d + 1}Z}\right
)^{-1/(2 - d)}
\end{equation}
Substituting this back into (\ref{varenergygend}) provides us with
dimensionless ground-state energy $\epsilon_{n} = E_{n}/|E_{1}|$ 
\begin{equation}
\label{gsenergygend}
\epsilon_{n} = -n\left (1 - {n -1 \over 2^{d + 1}Z}\right
)^{2/(2 - d)} 
\end{equation}

In one dimension Eqs.(\ref{varllgend}) and
(\ref{gsenergygend}) reduce to the results of Kadomtsev and Kudryavtsev
\cite{KK} obtained in the context of the ground-state properties of
ordinary atoms placed in a superstrong magnetic field. They are 
numerically just slightly less accurate than Eqs.(\ref{lld=1}) and
(\ref{gsenergyd=1}) and predict very similar physics.  Indeed, for $d
= 1$ the ground-state energy (\ref{gsenergygend}), $\epsilon_{n} =
-n[1 - (n - 1)/(2Z) + (n - 1)^{2}/(16Z^{2})]$, is only different from
(\ref{gsenergyd=1}) in the numerical coefficient of the last term
($1/16$ versus $1/12$). Moreover apart from the fact that the variational
treatment predicts $n_{max} = 4Z + 1$ (instead of $n_{max} = 2Z + 1$),
the divergences of the localization length, Eqs. (\ref{lld=1}) and
(\ref{varllgend}), near $n_{max}$ are identical. 

For general $d$ the result (\ref{varllgend}) can also be interpreted
using the language of a screening whose effect diminishes as the space
dimensionality grows.  Equivalently we can say that the higher
the dimensionality of space, the more particles can be bound.  This is
physically plausible as higher dimensionality implies more ``room''
around the nucleus.  The numerical accuracy of the variational predictions
degrades as the space dimensionality increases;  only for $d = 1$ does
the exponential dependence correctly solve the single-particle
problem.

For the marginal two-dimensional case when the potential energy
part of (\ref{varenergygend}) has the same $R^{-2}$ dependence as the
kinetic energy, variational predictions are not even qualitatively
correct:  for $n = 1$ ($\lambda = 0$) Eq.(\ref{varenergygend}) predicts a
threshold for the appearance of the bound state.  This contradicts the
known fact \cite{LL4} that there always is a bound state in a
two-dimensional radially symmetric potential well.

For $d > 2$ Eq.(\ref{varenergygend}) predicts a minimum as $R
\rightarrow 0$ which cannot be made consistent with the condition $R
\gg a$.

The reason why well-localized single-scale wave functions fail to
describe the Bose-atom in $d \ge 2$ lies in our assumption that we can
take $a = 0$.
Indeed, for $d \ge 2$ the true large-distance behavior of the bound-state 
single-particle wave function cannot be continued to the origin
without introducing divergences \cite{KS}.  Therefore in order to
recover the correct physics we need both to keep the range of the nuclear
potential $a$ finite and to improve the variational function.  

\section{Two-dimensional case}

In two dimensions the form of the trial wave function is suggested by
the solution of the single-particle problem \cite{LL4}:  we choose
$\psi(r) = A\ln(R/a)$ for $0 \le r < a$, $\psi(r)= A\ln(R/r)$ for $a
\le r < R$, and $\psi(r) = 0$ for $r \ge R$ where $A$ is normalization
constant.  With this choice the variational energy becomes
\begin{equation}
\label{varenergyd=2}
E(R) = {2\hbar^{2}n \over mR^{2}} \left [ \ln(R/a) - 2 \xi \ln^{2}(R/a)
+ 3\xi \lambda \right ]
\end{equation} 
Apart from logarithmic factors (which are expected in marginal 
dimensionality), Eq.(\ref{varenergyd=2}) has the same structure as
Eq.(\ref{varenergygend}) for $d = 2$.  

For $n = 1$ ($\lambda = 0$)
we find that the minimum of (\ref{varenergyd=2}) occurs for $R_{1}/a =
\exp[(1 + 2\xi +
\sqrt{1 + 4\xi^{2}})/4\xi]$.  The consistency condition $R_{1} \gg a$
becomes equivalent to the requirement of shallow nuclear well, $\xi
\ll 1$.  In this limit we find $R_{1} \simeq ae^{1/2\xi}$ thus
reproducing the known result \cite{LL4}.  The expression for the 
ground-state energy $E_{1} =
- (\hbar^{2}/mR_{1}^{2})(2\xi + \sqrt{1 + 4\xi^{2}})$ also reduces to
the correct answer \cite{LL4} $E_{1} \simeq
-(\hbar^{2}/ma^{2})e^{-1/\xi}$ in the $\xi \ll 1$ limit. 

As the trapping potential gets deeper, i. e. $\xi$ increases, the size of
the single-particle bound state $R_{1}$ monotonically decreases. In
the limit of very deep well we find $R_{1}/a = e[1 +
1/(4\xi)] > 1$ which meets expectation as for $\xi \gg 1$ the size of
the ground state must be of the order of the well size.  In a very
deep well the ground-state energy is expected to be greater than the
negative of the well depth (the classical answer) by the zero-point 
energy of a particle localized
inside the well, of order $\hbar^{2}/ma^{2}$.   Apart from a numerical factor of
order unity, the expression for the single-particle energy 
$E_{1} = - 2U_{0}/e^{2} + \hbar^{2}/ma^{2}e^{2}$ (specified
to the case of a rectangular well of depth $U_{0}$) conforms with the
physics.  Since this is a variational result, the exact ground-state
energy is lower which is reflected in the inequality $2/e^{2} < 1$.
We note that the localization length $R_{1}$ is at least $e$ times larger than
$a$, and binding more repelling particles can only increase the size of
the ground-state.  We conclude that the trial wave function which led 
to (\ref{varenergyd=2}) correctly captures the physics of the
two-dimensional Bose-atom for {\it any} value of $\xi$.  Below we
restrict ourselves to the case of shallow well, $\xi \ll 1$, when it
supports only one single-particle state.
\begin{figure}[htbp]
\epsfxsize=3.6in
\vspace*{-0.3cm}
\hspace*{-0.5cm}
\epsfbox{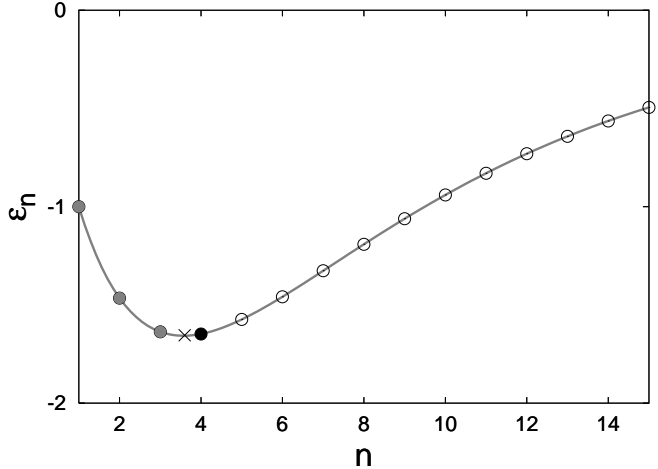}
\vspace*{0.1cm}
\caption{Dependence of the ground-state energy of a
two-dimensional Bose-atom on the number of bound particles $n$ for $Z =
1.5$ and $\xi = 0.1$.  The styling is the same as in Fig.4.}
\end{figure}
For general $n$ minimization of (\ref{varenergyd=2}) provides us with
the reduced size of the ground-state 
\begin{equation}
\label{lld=2}
\rho_{n} = R_{n}/R_{1} = \exp \left [{\sqrt{1 + 24\xi^{2}\lambda} -
1\over 4\xi} \right ], 
\end{equation}
where the $n$-dependence enters through $\lambda = (n - 1)/Z$,
Eq.(\ref{parameters}).  In contrast to the lower-dimensional
cases, $1 \le d < 2$, for finite values of $n$, $Z$, and $\xi \ll 1$,
the two-dimensional localization length (\ref{lld=2}) never diverges.
This implies that in two dimensions even a shallow well, 
$\xi \ll 1$, can bind an {\it arbitrary} number of bosons.  We note that
the size of the ground-state is very sensitive to the number of bound
particles.  For example, for $24\xi^{2}(n -1)/Z \ll 1$, Eq.(\ref{lld=2}) 
simplifies to $\rho_{n} \simeq \exp[3\xi(n - 1)/Z]$.  In this limit the
dependence on the nuclear strength effectively drops out as $v = \xi/Z$ 
parameterizes the interparticle repulsion alone [see
Eq.(\ref{parameters})].  In the opposite limit $24\xi^{2}(n -1)/Z \gg
1$, Eq.(\ref{lld=2}) reduces to $\rho_{n} \simeq \exp\sqrt{3(n -
1)/(2Z)}$, and the dependence on the nuclear strength $\xi$ drops out.  

The corresponding reduced ground-state energy, $\epsilon_{n} =
E_{n}/|E_{1}|$, is given by
\begin{equation}
\label{gsenergyd=2}
\epsilon_{n} = - n \sqrt{1 + 24\xi^{2}\lambda}\exp \left [{1 - \sqrt{1 + 
24\xi^{2}\lambda}\over 2\xi} \right ] 
\end{equation}   

The population dependence of the ground-state energy
(\ref{gsenergyd=2}) typical to weakly repulsive particles (not very
small $Z/\xi$) is sketched in Fig.7.

For $n \ge 1$ and not very strong interparticle repulsion, the function 
$\epsilon_{n}$ first decreases, while for large $n$ it asymptotically
approaches zero, thus implying that it has a minimum.  The minimum,
indicated in Fig.7 by the cross, provides us with the equilibrium
population of an atom strongly coupled
to the zero chemical potential bosonic reservoir.  The position of
this minimum can be found in closed form in the limit $24\xi^{2}(n
-1)/Z \ll 1$.  Then Eq.(\ref{gsenergyd=2}) simplifies to
$\epsilon_{n} \simeq -n\exp[-6\xi(n - 1)/Z]$ which has a minimum at $n
= n_{x}= Z/6\xi$.  Even though for fixed $Z$ and $\xi \ll 1$ we
find $n_{x} \gg 1$, it is straightforward to verify that the condition 
$24\xi^{2}(n
-1)/Z \ll 1$ still holds.   For the argument to work, $n_{x}$ must exceed
unity thus implying $Z > 6\xi$;  this tells us how weak the
interparticle interaction should be for the minimum to occur at $n >
1$. 

When the coupling to the reservoir is weak, the population of the
atom is quantized.  In Fig.7 the corresponding ground state is shown
by a solid circle.  The population
transitions take place whenever the degeneracy condition is met,
i. e. $\epsilon_{n} = \epsilon_{n-1}$; in the $24\xi^{2}(n
-1)/Z \ll 1$ limit this is solved by $n_{1} = (1 - e^{-6\xi/Z})^{-1} >
n_{x}$.  Similar to
what we found in the one-dimensional case, in the $n_{x} \gg 1$ limit
the difference $n_{1} - n_{x}$ approaches $1/2$.  This implies that in order to
understand the population transitions we can approximate the true
$\epsilon_{n}$ dependence by a parabola centered at $n =n_{x}$
thus arriving at a model similar to (\ref{hgsenergy}).

For fixed $\xi\ll 1$ as the nuclear charge $Z$ gets smaller, there will
be a value for which $n_{1} = 2$, i. e. the transition from the doubly
to singly occupied atom happens.  This takes place at $6\xi/Z = \ln2$;
for $Z < 6\xi/\ln2$ the interparticle repusions are too strong, and
only single-bosonic atom is stable. 

If a two-dimensional shallow well is used as a pipette to extract
particles out of the BEC, then the relevant quantity to look at is the
reduced ground-state energy, $\epsilon_{n}^{'} = \epsilon_{n} + (1 - \gamma )
n$.  For not very strong interparticle repulsions this function has a
minimum whose position can be experimentally tuned to extract
particles out of the BEC.

\section{Three-dimensional case}

In three dimensions the form of the trial wave function is also
suggested by the solution of the single-particle problem \cite{LL2}:
we select $\psi(r) = (A/a)e^{-a/R}$ for $0 \le r< a$ and $\psi(r) =
(A/r)e^{-r/R}$ for $r \ge a$ where $A$ is normalization constant.  Let
us first look at the single-particle case $n = 1$.  Substituting this
function into the HFGP functional (\ref{GP}) we arrive at
\begin{equation}
\label{varenergyn=1d=3}
E(R) = {\hbar^{2}\over ma^{2}}\left [ {8\xi - 1\over 6(R/a)^{2}} - {2\xi
- 1\over (R/a)}\right ]
\end{equation}  
This function has a minimum provided $\xi > \xi_{c} = 1/2$ which
corresponds to the well-known fact that in a three-dimensional well
the first bound state appears only for a sufficiently
deep well \cite{LL2}.  As an assesment of the accuracy of the
trial function, we note that for the exactly solvable case of
rectangular well of radius $a$ and depth $U_{0}$ \cite{LL2} the corresponding
$\xi_{c} = \pi^{2}/24 \simeq 0.4112$ is fairly close to the
variational threshold of $1/2$.  

Minimizing Eq.(\ref{varenergyn=1d=3}) for $\xi > \xi_{c} = 1/2$ we
arrive at the expressions for the localization length and the ground-state
energy, respectively
\begin{equation}
\label{llgsenergyd=3n=1} 
{R_{1}\over a} = {8 - \xi^{-1}\over 3(2 - \xi^{-1})}~~~~, ~~~~ E_{1} = -
{3\hbar^{2}\xi \over 2ma^{2}} {(2 - \xi^{-1})^{2}\over 8 - \xi^{-1}} 
\end{equation}
which define a system of units used below.

Requiring $R_{1} \gg a$ puts us in the vicinity of the critical value 
$\xi = \xi_{c}$.  Introducing the reduced distance from the threshold,
$\Delta = 1 - 1/(2\xi)$, and specifying to the $\Delta \ll 1$ limit
we find $R_{1} = a/\Delta$ and $E_{1} = - \hbar^{2}\Delta^{2}/2ma^{2}$ both
correctly reproducing the critical behavior near the threshold
\cite{LL2}.

For arbitrary $n$ the variational energy is given by
\begin{eqnarray}
\label{varenergyd=3}
E(R)& = & { \hbar^{2}n \xi \over ma^{2}} \Bigglb [{8 - \xi^{-1} + 16
\lambda \over 6(R/a)^{2}}\nonumber\\
&-& {2 - \xi^{-1} \over (R/a)} -  { 8 \lambda \ln(R/a) \over
(R/a)^{3}} \Biggrb ]
\end{eqnarray}  
where some of the $n$-dependence comes through $\lambda = (n -
1)/Z$, Eq.(\ref{parameters}). 
Even though the last term of (\ref{varenergyd=3}) is 
of higher order in $R/a \gg 1$, keeping it helps clarify the
physics of weak interparticle repulsions, as will become clear shortly.

The variational energy has a minimum whose existence is not affected
by the last term of (\ref{varenergyd=3}). Let us temporarily drop it
and look at the outcome.  Minimizing (\ref{varenergyd=3}) and using
$R_{1}$ from (\ref{llgsenergyd=3n=1}) as a unit of length we find
\begin{equation}
\label{lld=3}
\rho_{n} = R_{n}/R_{1} = 1 + Z_{c}(n - 1)/Z
\end{equation}
where we introduced a new parameter 
\begin{equation}
\label{criticalz}
Z_{c} = {16 \xi \over 8 \xi - 1}
\end{equation}
whose significance will become clear shortly. 

Similar to the two-dimensional case, for finite values of the
parameters the size of the ground-state never diverges,
thus implying that an arbitrary number of bosons can be bound.  The 
localization length (\ref{lld=3}) does not grow with $n$ as fast as its
two-dimensional counterpart (\ref{lld=2}) indicating a stronger ability
to bind particles than in two dimensions. In units of $E_{1}$ 
[Eq.(\ref{llgsenergyd=3n=1})], the reduced ground-state energy
$\epsilon_{n} = E_{n}/|E_{1}|$ is given by
\begin{equation}
\label{gsenergy1d=3}
\epsilon_{n} = - {n\over 1 + Z_{c}(n - 1)/Z}
\end{equation}
The interesting feature of this result is that the ground-state energy
has a finite negative limit of $-Z/Z_{c}$ as $n\rightarrow \infty$.
For $n \ge 1$ and $Z < Z_{c}$ the ground-state energy is a
monotonically increasing function of $n$ approaching the asymptotic
limit of $-Z/Z_{c}$ from below.  This implies that only the
single-bosonic atom is stable when the nuclear well is
weakly coupled to the zero chemical potential bosonic reservoir:  the 
interparticle repulsions are too strong and all the states with $n >
1$ are metastable.  As the reduced nuclear strength $\xi$ increases from
its threshold value of $1/2$ to infinity, the critical parameter
$Z_{c}$ varies insignificantly: it monotonically decreases from $8/3$ to $2$.

On the other hand, Eq.(\ref{gsenergy1d=3}) predicts that 
the ground-state energy is a monotonically decreasing function of $n$
for $Z > Z_{c}$, which approaches its asymptotic limit of $-Z/Z_{c}$
from above.  This would imply that the lowest energy state would have 
an infinite number of bound bosons.  We arrived at this conclusion
however neglecting the last negative term of
(\ref{varenergyd=3});  now we demonstrate that it actually introduces a
minimum in $\epsilon_{n}$ for $Z > Z_{c}$. 

Indeed substituting (\ref{lld=3}) in (\ref{varenergyd=3}) we find
a more accurate expression for the reduced ground-state energy
\begin{equation}
\label{gsenergy2d=3}
\epsilon_{n} = - {n\over \rho_{n}}
 -  {9Z_{c}^{2}\Delta n(n - 1)\ln(8\rho_{n}/3Z_{c}\Delta) 
\over 8Z \rho_{n}^{3}}
\end{equation}
where $\Delta = 1 - 1/(2\xi)$ is the dimensionless distance from the
threshold.  The second term of (\ref{gsenergy2d=3}) does not affect our earlier
conclusion that for $Z < Z_{c}$ the ground-state energy is a monotonically
increasing function of $n$ approaching $-Z/Z_{c}$ from below.

For $Z > Z_{c}$ Eq.(\ref{gsenergy2d=3}) has a minimum
whose position in the $n \gg 1$ limit is given by 
\begin{equation}
\label{atomcapacity}
n = n_{x} \simeq v^{-1}(2\xi - 1) \exp\left ({8\xi - 1 - 16v\over
18\xi - 9}\right )
\end{equation}
where $v = \xi/Z$ parametrizes interparticle repulsions, and we used 
$\Delta = 1 - 1/(2\xi)$ and the definition (\ref{criticalz}).  The 
ground-state energy evaluated at $n_{x}$ is lower than $-Z/Z_{c}$ and
for $n > n_{x}$ it monotonically increases approaching $-Z/Z_{c}$ from
below.  The minimum $n = n_{x}$ provides us with the average number of
bound bosons provided the nuclear well is strongly coupled to the zero
chemical potential bosonic reservoir.

In Fig.8 we sketch the dependence of the ground-state energy on the
number of bound bosons for the cases of strong ($Z < Z_{c}$, Fig.8a) and
weak ($Z > Z_{c}$, Fig.8b) interparticle interaction.

If the strength of the nuclear attraction $\xi > \xi_{c} = 1/2$ is kept
constant and the interparticle
repulsion goes to zero ($v \rightarrow 0$), or if the interparticle
repulsion $v$ is fixed while the depth of the trapping potential goes to
infinity ($\xi \rightarrow \infty$), the number of bound bosons
(\ref{atomcapacity}) increases to infinity.  These conclusions are in 
accordance with physical expectations.  
\begin{figure}[htbp]
\epsfxsize=3.6in
\vspace*{-0.3cm}
\hspace*{-0.5cm}
\epsfbox{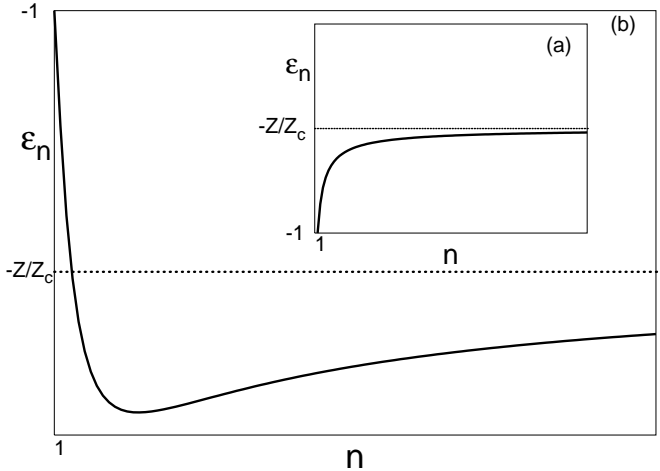}
\vspace*{0.1cm}
\caption{Ground-state energy of a three-dimensional Bose-atom as a
function of number of bound particles:  (a) strongly repulsive
particles ($Z < Z_{c}$), (b) weakly repulsive particles ($Z > Z_{c}$). }
\end{figure}
On the other hand, the limit of fixed interparticle repulsion and 
nearly-threshold or resonant nuclear potential, 
$\Delta = 1 - 1/(2\xi) \ll 1$, is more surprising.  To
better appreciate how unusual the behavior is we note that in the
vicinity of the single-particle threshold $\xi = \xi_{c} = 1/2$ the
number of bound bosons $n_{x} \simeq (\Delta/v)\exp[(3 -
16v)/(9\Delta)]$ {\it increases} as the nuclear well gets 
{\it less attractive}, $\Delta \rightarrow 0$; at $\Delta = +0$ the
number of bound particles is {\it infinite}.  In the same limit, the
magnitude of the many-body
ground-state energy $E = - 3\hbar^{2}\Delta^{2}/32ma^{2}v$ {\it
decreases} while the size of the atom {\it diverges} exponentially, $R
\simeq a\exp[(3 - 16v)/(9\Delta)]$.

Qualitatively this effect of enhanced binding due to resonant nuclear
potential can be explained by noticing that upon approach the
threshold, $\Delta \rightarrow 0$, the size of the 
single-particle bound state, $R_{1} = a/\Delta$, increases without
bound.   This means that the volume wherein the nuclear attraction is
felt {\it increases} thus making it possible to accomodate more bosons -
nonresonant short-range interparticle repulsion becomes progressively
less important and cannot overcome the energy gain of condensation as
$\Delta \rightarrow 0$.

In one and two dimensions a single-particle bound state exists for
arbitrary weak trapping potential thus implying that $\xi_{c} = 0$.
However one- and two-dimensional Bose-atoms do not exhibit enhanced
resonant binding as despite the divergence of the localization length as 
$\xi \rightarrow 0$ there is not enough ``room'' around the
nucleus to accomodate repulsive bosons: interparticle
repulsions play a far more important role in low dimensions.  Moreover
for $\xi = 0$ and $d \le 2$ there are only repulsive interactions in
the system and no bound state can form.  On the other hand, for $d =
3$ and $0 < \xi \le \xi_{c}$ there are both repulsive and attractive 
interactions in the system, and a bound state is not prohibited on
general grounds.

The effect of enhanced binding due to resonant interaction with an
attractive center bears qualitative similarity with the Efimov
effect occuring in a system of three bosons \cite{Efimov1} where
resonant two-body forces trigger a formation of an arbitrarily large number of
loosely bound levels in a three-particle system.  Specifically,
the Efimov states occur due to an effective long-range interparticle 
attraction induced by the proximity of the two-body resonance.  This 
interaction, decaying as an inverse square of distance, is
universal in a sense that it does not depend on the shape of the
two-body potential;  both at small and large distances it is cut off by the
range of two-body forces and the size of the two-particle bound state,
respectively.  

Formally our effect occurs due to the last negative term of
Eq.(\ref{varenergyd=3}), and the fact that it is proportional to the
number of pair interactions $n(n-1)/2$ tempts us to interpret it as
originating from an effective long-range {\it attraction} of the
$\hbar^{2}a_{s}\ln(r/a)/mr^{3}$ form where we expressed $g$ in terms
of the low-energy $s$-wave scattering length $a_{s}$ \cite{DGP,GP}: $g =
4\pi\hbar^{2}a_{s}/m$.  This interpretation is not entirely
implausible as the particle attraction to the resonant nuclear well,
occuring at distances significantly exceeding the scattering length
$a_{s}$, also pulls the bosons towards each other thus inducing an
effective long-range attraction.  Similar to the Efimov effect, the 
$\hbar^{2}a_{s}\ln(r/a)/mr^{3}$ interaction is universal:  it
has its origin in the large-distance behavior of the wave
function and it is characterized by a very weak (logarithmic)
dependence on the size of the nuclear well $a$.  At large distances
the induced interaction is cut off by the size of the many-body state
while at small distances it vanishes at $r \simeq a$.   

We note that our effective interaction is different from the Efimov
$1/r^{2}$ form.  The difference will not appear that surprising if
we look at two identical repulsive bosons attracted to a much heavier 
particle mimicking the nucleous of our problem.  It turns out
that if the mass of the heavy particle approaches infinity, the 
effective $1/r^{2}$ attraction and thus the Efimov effect
disappear \cite{Efimov1}.   This makes us think that our effect of enhanced
binding to the resonant nucleous is {\it not} a manifestation of the
original Efimov effect in a many-body system.  It is curious that for
the three-boson problem the correction to the $1/r^{2}$ Efimov attraction is
also universal and has the $\hbar^{2}a_{s}/mr^{3}$ form \cite{Efimov2}.  Apart
from logarithmic factor this is what we find.  

The arguments of the last two paragraphs are no more than suggestive - the
translationally invariant three-body problem is somewhat different
from that of the $n$-boson atom where the translational symmetry is
broken, and complementary work is necessary to better understand the
physics of our effect.     
\begin{figure}[htbp]
\epsfxsize=3.6in
\vspace*{-0.3cm}
\hspace*{-0.5cm}
\epsfbox{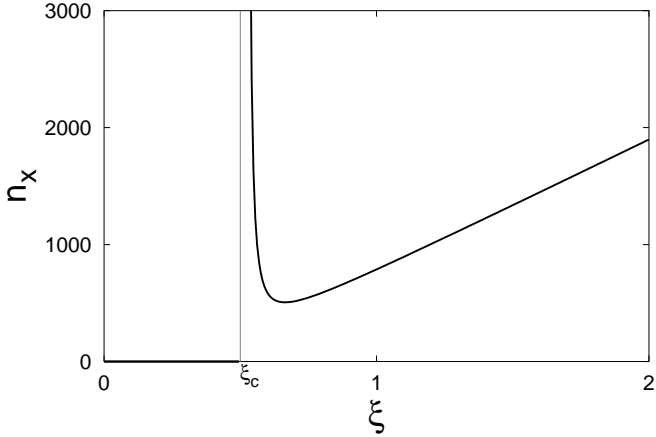}
\vspace*{0.1cm}
\caption{Equilibrium number of bound particles in a three-dimensional 
Bose-atom as a function of dimensionless depth of the nuclear
potential $\xi$ for fixed interparticle repulsion satisfying $Z >
Z_{c}$ or equivalently $8\xi - 1 - 16v > 0$; $v = 0.00275$ was used to
construct the graph. }
\end{figure}
In Fig.9 we sketch the dependence of the number of bound bosons on the
dimensionless nuclear strength $\xi$, given by Eq.(\ref{atomcapacity}),
for the case of sufficiently weak ($Z > Z_{c}$) interparticle
repulsion.  The enhanced resonant binding corresponds to unlimited growth of
$n_{x}$ as $\xi$ approaches $\xi_{c} = 1/2$ from above. 

In the limit of weak coupling to the zero chemical potential bosonic
reservoir, the population of the atom is quantized and first-order
population transitions take place whenever the condition of degeneracy 
holds ($\epsilon_{n} = \epsilon_{n - 1}$).  In the limit $n_{x} \gg 1$
when the $\epsilon_{n}$ dependence can be approximated by a parabola 
centered at $n = n_{x}$, the transitions happen at half-integer $n_{x}$.   

When a three-dimensional trapping potential is used as a pipette to
extract particles out of the BEC,  the single-particle ground state remains 
unoccupied as long as it is higher in energy than the BEC chemical potential.
This effectively cuts off the resonant binding singularity at 
$\xi = \xi_{c}$.   As a result the $\xi$ dependence of the equilibrium
pipette population may or may not have a segment where the particle
number decreases as the pipette potential gets deeper.
Below we ignore this possibility because the resonant region is rather
narrow, due to the exponential 
dependence in (\ref{atomcapacity}) .    The reduced ground-state energy, 
$\epsilon_{n}^{'} = \epsilon_{n} + (1 - \gamma)n$, is given by:
\begin{equation}
\label{gsenergy'd=3}
\epsilon_{n}^{'} = - {n\over 1 + Z_{c}(n - 1)/Z} + (1 - \gamma ) n
\end{equation} 
Since adding the $(1 - \gamma ) n$ term to (\ref{gsenergy1d=3}) inevitably
introduces a minimum in $\epsilon_{n}^{'}$, we can ignore the last term
of (\ref{gsenergy2d=3}).  It amounts to the assumption that the
minimum of (\ref{gsenergy2d=3}) lies at a substantially larger value of
$n$ than that of (\ref{gsenergy'd=3}) which can be always accomplished.  

The ground-state energy (\ref{gsenergy'd=3}) has a minimum at
\begin{equation}
\label{capacity}
n = n_{x} = 1 + (Z/Z_{c})\left [\sqrt{[1 - Z_{c}/Z]/(1 - \gamma )} - 1 \right ]
\end{equation}
whose position can be experimentally tuned to extract particles out of
the BEC.  In the vicinity of $n_{x}$ the energy (\ref{gsenergy'd=3}) can
be represented as $\epsilon_{n}^{'} = \epsilon_{n_{x}}^{'} + Z_{c}(1 - \gamma )^{3/2}(n
- n_{x})^{2}/Z\sqrt{1 - {Z_{c}\over Z}}$.  In the original units this
introduces an energy scale
\begin{equation}
\label{chenergy}
E_{c} = |E_{1}|{Z_{c}(1 - \gamma )^{3/2}\over Z\sqrt{1 - {Z_{c}\over Z}} }
\end{equation}
Since the divergence of $E_{c}$ at $Z = Z_{c}$ corresponds to
$n_{x} = 0$, it does not reflect any physical singularity.
The energy scale (\ref{chenergy}) parallels the charging energy of 
electronic systems \cite{Likharev} and should exceed the temperature
$T$ for the single-particle effects to be observed.  In the limit of weak
interactions ($Z \gg Z_{c}$) when $E_{c} \sim Z_{c}/Z$ and $n_{x} \gg 1$ the
condition $E_{c} > T$ inevitably breaks down -  finite condensate
temperature will make it impossible to control precisely a very large
number of particles.  On the other hand as $Z$ approaches $Z = Z_{c}$
from above, the inequality $E_{c} > T$ can be satisfied, implying
that a precise few-particle manipulation is realizable in
practice despite the finite condensate temperature.

\section{Rounding of the boson staircase by tunneling}

In the discussion above we ignored particle discreteness 
in the regime of strong tunneling, while for
weak tunneling the quantization of the atom (pipette) population 
played an important role.  Here we
develop a quantitative theory of weak finite tunneling which
provides a link  between the two limiting cases. 

In the preceding discussion we assumed that the total number of 
particles in the system and the number in the pipette are both 
good quantum numbers.   The energy spectrum for the system was
derived from the spectra for the condensate reservoir and the pipette
potential, treated as separate problems.  Tunneling couples states 
of the system having a fixed total number of 
particles and differing numbers in the pipette.  This is a small
effect that can be treated perturbatively when the states 
of different $n$ have significantly different
energies, but becomes important in case of degeneracy.  The discussion
of the separated systems shows that we need consider only the
case of a simple degeneracy of states with $n-1$ and $n$ particles 
($E_{n-1}^{'} = E_{n}^{'}$),
and then all other population states can be ignored.  
Let these two states be coupled by tunneling with a matrix element
$M$.  The Hamiltonian of the system reduces to a $2 \times 2$ matrix
$H_{ik}$ where the diagonal entries $H_{11} = E_{n - 1}^{'}$, $H_{22}
= E_{n}^{'}$ are the previously computed ground-state energies and the 
off-diagonal element $H_{12} = H_{21} = M$ is estimated below.  The
matrix elements all depend on the external parameter
$n_{x}$ and as it changes slowly from the region where $E_{n -
1}^{'} < E_{n}^{'}$ to $E_{n - 1}^{'} > E_{n}^{'}$, the $n - 1$
particle state adiabatically evolves into an $n$ particle state.  

The Hamiltonian $H_{ik}$ has eigenvalues 
\begin{equation}
\label{secular}
E^{'} = (E_{n - 1}^{'} + E_{n}^{'})/2 \pm \sqrt{(E_{n - 1}^{'} -
E_{n}^{'})^{2}/4 + M^{2}}
\end{equation}
which show that finite tunneling replaces the crossing of the energy
levels by the avoided crossing.   At the former crossing point, 
$E_{n - 1}^{'} = E_{n}^{'}$, there is an energy gap of magnitude $2M$
between the ground state (the lower root) and the first excited state 
(the upper root).  Fig. 2 of the Introduction is the illustration of 
Eq.(\ref{secular}) where we chose $E_{n}^{'} \sim (n - n_{x})^{2}$ and
then periodically continued the graph.

The average population of the pipette can be found as $<n> = (n -
1)a_{1}^{2} + na_{2}^2$ where $a_{1}$ and $a_{2}$ are the probability 
amplitudes to be in the $n - 1$ and $n$ particle states, respectively.  
The amplitudes $a_{1,2}$ are given by the components of the normalized
eigenvector corresponding to the ground-state eigenvalue in
(\ref{secular}).  The result of this calculation is
\begin{equation}
\label{<n>} 
<n> = {n - 1 + n(\delta + \sqrt{\delta^{2} + 1})^{2} \over 2(\delta^{2} +
\delta \sqrt{\delta^{2} + 1} + 1)}
\end{equation}
where $\delta = (E_{n - 1}^{'} - E_{n}^{'} )/2M$ measures distance
between unperturbed energy curves in units of the energy gap $2M$.
The $\delta$-dependence of the equilibrium pipette population $<n>$
(\ref{<n>}) has a form of a rounded step which interpolates between $n - 1$ 
and $n$ as $\delta \rightarrow \mp \infty$, respectively, and has 
inversion symmetry about the step center $(0; n - 1/2)$.  
This implies that for small $M$ except near the crossing 
$E_{n - 1}^{'} = E_{n}^{'}$, the average population of the pipette is 
either nearly $n - 1$ ($E_{n - 1}^{'} <
E_{n}^{'}$) or nearly $n$ ($E_{n - 1}^{'} > E_{n}^{'}$), i. e. there
is very little mixing of the two states.   At the crossing 
$E_{n - 1}^{'} = E_{n}^{'}$ the average population is always half-integer 
$n - 1/2$ as expected.  As the tunneling matrix element $M$ gets larger,
the mixing between the two states affects a larger parameter region
around the crossing $E_{n - 1}^{'} = E_{n}^{'}$ which results in
progressively rounded staircases.   

Using the two-state approximation to trace a range of $n_{x}$ (as is
done in Fig. 2) will lead to small anomalies (quadratic in $M$) at 
the points where we 
change point of view on which pair of states are being mixed; this
minor defect could be removed by enlarging the matrix $H_{ik}$ to 
include more values of $n$.  
However, the two-state approximation is entirely adequate in
describing the physically interesting region of the vicinity of the 
level crossing  $E_{n - 1}^{'} = E_{n}^{'}$ where we find $<n> - (n -
1/2) = (E_{n - 1}^{'} - E_{n}^{'})/4M$.  In previous Sections it was 
demonstrated that if one is only concerned with the population
transitions, then for not very strong interparticle repulsion the
energy curves can be approximated by $E_{n}^{'} = const + E_{c}(n -
n_{x})^{2}/2$.  With this choice we find $<n> - (n - 1/2) =
(E_{c}/4M)[n_{x} - (n - 1/2)]$, i. e. the slope $d<n>/dn_{x}$ of the
$<n>$ on $n_{x}$ dependence at the degeneracy point $n_{x} = n - 1/2$ 
is $E_{c}/4M$. 

The system will remain in the lowest energy state if the parameter
$n_{x}$ is varied sufficiently slowly.  In an experimental realization,
the pipette parameters will be changed on a finite time scale $\tau$.  
According to the Landau-Zener theory \cite{LL1}, the adiabatic
behavior will be observed provided that the energy gap $E_{g}$
satisfies the condition $E_{g} \gg \hbar / \tau$. Thus it becomes
important to have an estimate of the size of the energy gap.
According to Eq.(\ref{secular}) this is determined by the
the matrix element of the trap potential that mixes the pipette state
with the condensate wavefunction
\begin{eqnarray}
\label{gap}
M& =& \int \psi_{pipette}({\bf  r} ) |U({\bf r} )| \psi_{BEC}({\bf
r} ) dV\nonumber\\
&\simeq& U_{0}a^{3}\psi_{pipette}({\bf  r}_{p} )\psi_{BEC}({\bf
r}_{p} )
\end{eqnarray}
where the wave functions are normalized at unity.  Since the pipette
is located in the classically forbidden region, the magnitude of the
energy gap will depend strongly on the distance between the
pipette and the edge of the condensate droplet.  
The discussion of Section VI implies that the pipette wave
function evaluated at the pipette location can be estimated as
$\psi_{pipette}({\bf r}_{p}) \simeq 1/(a\sqrt{R_{n}})$ where $R_{n}$
is the localization length of the pipette ground state.

\section{Experimental feasibility}

Below we address some issues of practical matter, namely, we explain
how to build a pipette suitable for single-particle manipulation, and
discuss experimental limitations set by nonzero temperature of the
condensate and finite time of tunneling.   We restrict ourselves 
to exploring the feasibility of observing single-particle effects 
in three-dimensional systems.

\subsection{Trap implementation} 

Very small atom traps can be created using optical dipole forces
\cite{Grimm00}, in which atoms are attracted to the intense region of
a focused laser beam.  Tight three-dimensional confinement has been
demonstrated at the intersection of two beams, each with a diameter of
about 15 $\mu$m \cite{Adams95,Ido00,Takasu03}.  However, for a trap of
this size, the zero-point energy $\hbar^2/ma^2$ is below
1 nK which is lower than achievable BEC temperatures.  This implies that
the finite temperature of the condensate is sufficient for excitation away
from the single-particle ground state.  Tighter
confinement could be achieved with stronger focusing of the beams. 
In principle, beam sizes comparable to the laser
wavelength can be achieved, with typical wavelengths
around 1 $\mu$m. At this value,
the zero-point energy is about 20 nK which is within the range of 
observed condensate temperures \cite{Cornish00,Weber03}.  Unfortunately,
achieving such tight focussing is very difficult.
Alternative approaches can be considered,
including holographic techniques \cite{Newell02,McGloin03},
optical superlattices \cite{Wasik97,Gorlitz01}, and
near-field techniques \cite{Shin03}.  

We also describe a novel approach which may be
particularly well suited for the atom pipette \cite{sackett03}.  Rather than
using a laser, an optical trap can be created at an interference fringe
produced by a broadband lamp which has been optically filtered to
remove wavelengths blue of the principal atomic transition.  If light
from the lamp is split into two
beams and the beams directed to be counterpropagating, 
then each frequency component creates an
independent optical standing wave.  Ordinarily the effect of the different
standing waves averages out, but at precisely the point where both
beams have travelled the same distance from the splitter, the standing
waves have a uniform phase and an observable interference pattern is 
produced.  
For a source with frequency bandwidth $\Delta\nu$ the interference
will extend for a distance $c/\Delta\nu$, so by using a broad source,
a narrow fringe can be obtained.  We estimate that trap sizes
below 1 $\mu$m can be achieved using light from a low-power
Hg arc lamp, with trap depths of up to several $\mu$K.

We assume, then, that a method such as these is used to 
generate a trap with $a \simeq 1$ $\mu$m, which is adjacent to a 
BEC of $^{23}$Na atoms.  Using the 
definition of the pseudopotential 
$g = 4\pi\hbar^{2}a_{s}/m$ \cite{DGP,GP} and Eq.(\ref{parameters}) the 
reduced interaction $v = \xi/Z$ parametrizing the interparticle repulsion 
turns into $v = a_{s}/a$.  For $^{23}Na$ particles we have $a_{s} = 2.75nm$
\cite{DGP} which leads to $v = 0.00275$;  this is the value used to
produce Fig. 9.  In principle the strong interaction regime $v \simeq
1$ can be achieved by applying a magnetic field, which increases the
scattering length via bringing the system in the vicinity of the 
Feschbach resonance \cite{note3}  

\subsection{Finiteness of the BEC temperature}

One of the main practical obstacles to the realization of the effects
we have predicted is the finite BEC temperature which, to be definite, 
will be set at $4nK$. 

The zero-point energy $\hbar^{2}/ma^{2}$ of a localized $^{23}Na$ atom is of
order $20nK$ which is sufficient to neglect its thermal agitation.  The 
maximal depth of our trap will be assumed to be $1\mu K$, thus
implying $\xi = 50$.  The minimal depth is determined from the
condition that the magnitude of
the single-particle energy $E_{1}$, Eq.(\ref{llgsenergyd=3n=1})
exceeds the BEC temperature;  this leads to $\xi \simeq 1$.  Therefore
by changing the laser power so that the reduced
nuclear strength $\xi$ varies between $1$ and $50$, the finiteness of
the BEC temperature can be made negligible as far as single-particle
excitations are concerned.  On the other hand, by looking at Fig. 9 
we conclude that finite BEC temperature prevents us from seeing the 
enhanced resonant binding which is limited to the vicinity of the 
single-particle threshold.  

If a trap is used as a pipette, then to observe single-particle effects we
also need to make sure that the change of the system energy upon adding or
removal of one particle is larger than the temperature.  This is equivalent
to requiring that the ``charging'' energy (\ref{chenergy}) exceeds the 
BEC temperature.  For $v = 0.00275$ and $\xi$ varying between $1$ and
$50$, we find that as the laser power goes up, the reduced nuclear
charge $Z = \xi/v$ increases from about $350$ to $18000$.  Substituting
these values in Eq.(\ref{chenergy}) and using the expression for the
single-particle energy from (\ref{llgsenergyd=3n=1}) we fnd that
for $\gamma \simeq 0$ the charging energy is an order of
magnitude smaller than the temperature.  However the inequality 
$E_{c} > T$ can be achieved by magnetically increasing the
scattering length.

\subsection{Adiabaticity}

Another aspect that needs to be considered is adiabaticity.  Tunnelling
of particles from the condensate mixes states with different numbers
of particles in the pipette; instead of there being a degeneracy for
certain values of $n_{x}$ there is an energy gap $E_{g}$.  

In order to estimate the condensate wave function at the pipette
location we assume that the condensate is confined by a harmonic
potential of frequency $\omega$.  Outside the BEC the nonlinearity of
the corresponding Gross-Pitaevskii equation \cite{DGP,GP} is
negligible so that in the classically forbidden region the wave
function will have the form of a harmonic oscillator 
function.  The function we should choose corresponds to the $l$th
excited state where $l$ is determined by the BEC
chemical potential, $\mu  = \hbar\omega (l+ 1/2)$.  In other words it is a
product of the Gaussian, 
$\exp[-r^{2}/(2a_{ho}^{2})]$, and $l$th Hermite function where 
$a_{ho} = [\hbar/(m\omega)]^{1/2}$ is the harmonic oscillator 
length \cite{DGP}.  This behavior is only valid outside the condensate 
droplet - as we move in, instead of all of the oscillations of the
$l$th Hermite polynomial we will get just the slow variation indicated by the 
Thomas-Fermi result \cite{DGP}.  Outside the condensate, the $l$th 
Hermite function is dominated by its highest term  $r^{l}$, so that we
find $\psi_{BEC} \propto r^{l}\exp[-r^{2}/(2a_{ho}^{2})]$ where the
exponent $l$ depends on the number of the condensate particles $N$.
Since this is a solution to the linear differential equation, the
overall prefactor is not determined by it, and this requires matching
the behavior inside the BEC to that outside.  At the classical edge of
the BEC droplet the wave function should satisfy $\psi_{BEC}(R_{TF})
\simeq 1/R_{TF}^{3/2}$ where $R_{TF} = a_{ho}(15Na_{s}/a_{ho})^{1/5}$ 
is the Thomas-Fermi radius of the condensate \cite{DGP}.  The 
harmonic-oscillator-type function that agrees with this has the form        
\begin{equation}
\psi_{BEC}({\bf r}) \simeq R_{TF}^{-3/2 - l}r^{l}\exp[-(r^{2} - R_{TF}^{2})/(2a_{ho}^{2})]
\end{equation}
Evaluation of this function at the pipette location $r_{p} = R_{TF} +
D$ leads to $\psi_{BEC}({\bf r}_{p}) \simeq
R_{TF}^{-3/2}\exp(-DR_{TF}/a_{ho}^{2})$ where we used the fact that
the distance between the BEC edge and the pipette $D$ is
significantly smaller than the Thomas-Fermi radius of the condensate,
$R_{TF}$.  This result together with $\psi_{pipette}({\bf r}_{p}) \simeq
1/(a\sqrt{R_{n}})$ provides us with the estimate for the energy
gap $E_{g} = 2M$, Eq.(\ref{gap}):
\begin{equation}
\label{estimate}
E_{g} \simeq U_{0} (a/R_{TF})^{3/2}(a/R_{n})^{1/2} \exp( - D/L)
\end{equation}
where the length scale is $L = a_{ho}^{2}/R_{TF}$.  It is desirable to 
have $L$ fairly short, so that the pipette will rapidly decouple from 
the condensate droplet as it is moved away; however, it should also be 
large enough that there is a significant tunnelling rate when the
pipette is at closest approach.   For $a_{ho} = 10^{-5} m$ and $R_{TF}
= 10^{-4} m$ we find $L = 10^{-6} m$.  With the choices $U_{0} = 1 \mu
K$ and $a = 10^{-6} m$, this requires only that the time scale over
which the pipette parameters are changed satisfies 
$\tau > 10^{-2}(R_{n}/a)^{1/2}\exp(D/L)$ seconds.  This is not
very restrictive unless the number of particles in the pipette is too
large or $D \gg L$.

\section{ACKNOWLEDGMENTS}  

We are grateful to C. A. Sackett for writing Section VIIIA of the paper
and to M. Timmins for his help in preparing the Figures.  This work
was supported by the Thomas F. Jeffress and Kate Miller Jeffress 
Memorial Trust, and by the Chemical Sciences, Geosciences and 
Biosciences Division, Office of Basic Energy Sciences, Office of Science, U. S.
Department of Energy.  Ryan M. Kalas was supported in part by a
Research Corporation Research Innovation Award.


\begin{references}
\bibitem{note3}  In a recent experiment on $^{85}Rb$ condensates
\protect\cite{Cornish00}, a four orders of
magnitude increase of $a_{s}$ was observed in the vicinity of the
Feschbach resonance.
\end{references}
\end{document}